Examination of an artifice for propelling a ship by the principle of internal motion
which was once proposed by the most astute man, Jacob Bernoulli
Author
L. Euler

E137-- *Examen artificii navis a principio motus interno propellendi quod quondam ab acutissimo viro Iacobo Bernoulli est propositum*
*Auctore*
*L. Eulero*

Translated and Commented by ChatGPT
Supervised by Sylvio Bistafa
September 2024

Foreword

The following translation of Leonhard Euler's "Examination of an Artifice for Propelling a Ship by the Principle of Internal Motion," originally published in 1750, offers a glimpse into a fascinating historical debate in the field of mechanics. This work critically examines a proposal by Jacob Bernoulli, one of the foremost mathematicians of the 17th century, who suggested that a ship could be propelled by forces entirely confined within the vessel itself—a concept that contradicts the principles of classical mechanics.

Euler, whose contributions to mathematics and physics are unparalleled, approaches Bernoulli's proposition with both respect and skepticism. Through meticulous analysis, Euler demonstrates that the forces generated by internal mechanisms, such as Bernoulli's pendulum, cannot impart any net motion to the ship. Euler's work is a testament to the rigorous application of Newtonian mechanics, emphasizing the importance of external reference points in the generation of motion.

This translation is intended to make Euler's insights accessible to a modern audience, highlighting the enduring relevance of these foundational principles. The commentary provided offers context and clarification, drawing parallels between Euler's work and contemporary understandings of mechanics. By examining the arguments of two of the greatest minds in the history of science, readers are invited to appreciate the depth and precision of classical mechanics as it was developed during the Enlightenment.

This translation was prepared using the assistance of advanced language models to ensure accuracy and clarity, with the supervision and expertise of Sylvio Bistafa. It is our hope that this work will contribute to a greater understanding of the historical development of mechanics and the ongoing dialogue between past and present scientific ideas.

§. I.

In the works of Jacob Bernoulli, which were published last year in Geneva, page 1109, there is a note inserted with the following title: "The device for propelling a ship by internal motion confined within the ship itself," in which the illustrious man tries to show that, although it is commonly thought that ships can only be propelled by forces applied externally, it is nevertheless possible for a ship to be set in motion by internal force alone. Though this device may appear highly paradoxical, if it could indeed produce the desired effect, it would deservedly be preferred over most other methods by which ships are commonly propelled. Since it is not certain whether an experiment has ever been made or whether it succeeded as hoped, it seems worthwhile to examine this mechanism more closely and evaluate it according to the laws of motion.

This passage refers to an early concept or proposition by Jacob Bernoulli regarding the idea of using internal motion to propel a ship.

§. II. When a sailor standing on firm ground can propel a ship with a pole but cannot achieve the same while standing on the ship itself, this is because, as much as he pushes the ship forward, he presses the keel backward with his feet by the same amount. Indeed, it seems correctly concluded that it is not possible to impart motion by a force that exists entirely within the ship. For no matter how much men or other machines placed in the ship strive to propel it, since the reaction is always equal to the action and both are sustained equally by the ship, no motion is thereby achieved. Hence, all efforts by those aboard the ship to propel it are in vain, unless they are able to apply themselves to the shore or another body situated outside the ship.

This passage elaborates on the concept that internal forces within a ship cannot generate motion, emphasizing the necessity of an external reference point for propulsion.

§.3. Bernoulli was by no means ignorant of this truth, but he did not believe it applied to all kinds of forces. He thought it should be restricted only to those forces that are commonly called *dead*, which are contained solely in pressures. However, another kind of force, called *living* forces, which arise from impact, he believed to be exempt from this law. Hence, he did not doubt that in a ship, such impacts and collisions could be produced that would induce motion in the ship. This opinion, if explained according to the views of most modern philosophers, who establish a sharp distinction between *living* and *dead* forces, might seem to rest on a very firm foundation. However, since I have shown that this distinction is without any foundation, and that nothing can be accomplished by *living* forces that cannot also be achieved by *dead* forces, it is highly to be feared that all the motion which Bernoulli sought to impart to ships by means of impacts will come to nothing.

This passage discusses Bernoulli's distinction between "living" and "dead" forces and challenges the validity of this distinction, suggesting that Bernoulli's approach to generating motion in ships might ultimately be ineffective.

§.4. The machine that Jacob Bernoulli proposed for this purpose is as follows: He orders a solid platform **AF** to be installed in the ship **DEFG**, positioned perpendicular to the horizon. This platform should be perfectly elastic, made of steel or some

reticulated material, especially at point **C**, where it receives the impact. On this platform, at point **A**, is suspended a pendulum **AB** with an attached weight **B**, also perfectly elastic. As it rises through the arc **BC**, it strikes the platform, and at the same time, propels the entire ship forward toward the bow **G**. After the impact, due to its elasticity, it rebounds and, by descending again, repeats the cycle of impacts continuously, thus imparting perpetual motion to the ship. To prevent the motion of the pendulum from gradually weakening due to air resistance, and to ensure that the pendulum consistently rises to the starting point of its arc **B**, he indicates that this can be achieved with the help of an automatic mechanism, as is done in pendulum clocks.

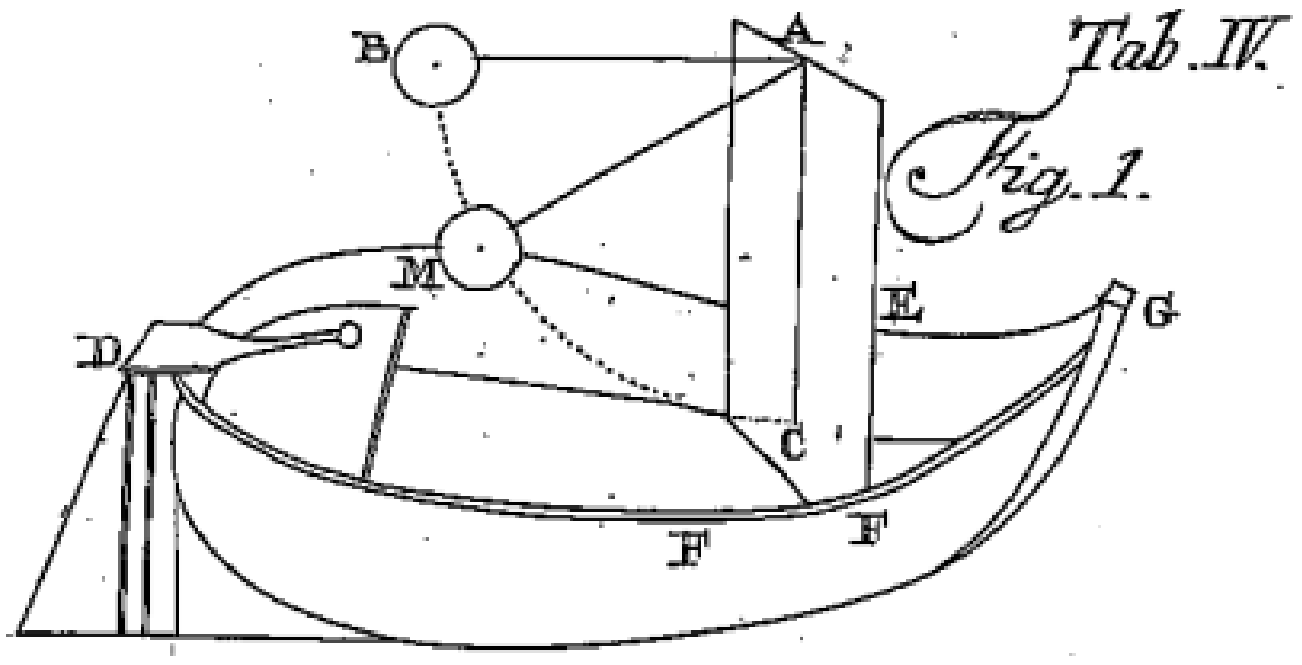

This passage describes the mechanical setup that Bernoulli proposed, utilizing a pendulum to induce continuous motion in the ship.

§.5. If we consider only the successive impacts by which the platform **AF** is continuously struck, there is absolutely no doubt that these would propel the ship into motion, and the mass of the pendulum could easily be increased to the point where, after overcoming the resistance of the water, the ship would attain a significant speed, no matter how large it is. However, it must also be noted that, as the pendulum alternately ascends and descends, it exerts an opposing force on the ship, pulling it toward the stern **D**. For, at any position **AM**, the pendulum is acted upon both by its weight and by the centrifugal force, which is sustained by the suspension point **A**, and pulls the pendulum in the direction of **AM**. Since this force is always directed backward, it pushes the ship backward. Hence, the ship's propulsion toward the bow **G** will only result if the forces of impact exceed these continuous backward-directed forces. If this excess can be achieved, the ship will move forward.

This section discusses the opposing forces caused by the pendulum's motion and how the ship's forward movement depends on the pendulum's impact force overcoming the backward pull.

§.6. Most philosophers, who follow Leibniz's ideas on forces but have misunderstood them, and who believe *living* forces to be infinitely greater than *dead* ones, will undoubtedly assert that the ship will achieve significant motion in this manner, and they will not consider it necessary to account for the forces that would push the ship backward, as the propelling force of the impacts seems to them incomparably greater. However, the very astute Jacob Bernoulli thought quite differently. He carefully investigated the effect of these *dead* forces and subtracted them from the effect produced by each impact in order to arrive at the true propulsion of the ship. He found through calculation that the forces from the impacts somewhat exceed the forces that

continuously pull the pendulum backward, and from this he finally concluded that the ship should indeed be propelled by such a pendulum.

This section contrasts the views of philosophers who overestimate the difference between *living* and *dead* forces with Bernoulli's more careful and precise approach, emphasizing his mathematical calculations in determining that the ship could indeed be propelled by the pendulum.

§.7. Although he found that the force by which the ship is propelled by the impacts of the pendulum is not much greater than the opposing force arising from tension, he nevertheless believed that this force would impart a considerable motion to the ship. Even when taking into account the resistance of the water, he estimated that the ship could attain a notable speed. In a scenario where the weight of the pendulum is assumed to be equal to one-hundredth of the total weight of the ship, he calculated that the ship should be propelled a distance of 82½ feet per minute, without considering any reduction in resistance due to an appropriate design of the prow. When this case is adapted to a ship with a beaked prow, for which he assumed the resistance to be ten times smaller, he concluded that the speed imparted to the ship could exceed 260 feet per minute, or 15,649 feet per hour. This speed is certainly so great that it would scarcely be exceeded by conventional rowing.

This section shows Bernoulli's calculations of the ship's potential speed when propelled by the pendulum, suggesting that it could rival traditional methods of propulsion like rowing.

§.8. If this method of propelling ships were as effective as described, there would be no doubt that it would not only be far superior to rowing but could also, in many cases, be employed with great success when there is a lack of wind. For, considering the mass of the ship, a significant amount of force is usually required to move the oars, whereas in this mechanical system, almost no force is needed. Once the pendulum has been elevated to its highest position, after the first impact, it ascends almost to the same height on its own due to the great elasticity of both the pendulum itself and the platform. Whatever ascent is lost during each oscillation due to air resistance or the lack of perfect elasticity can be easily restored with a small amount of force, so much so that the continuous motion of the pendulum could be maintained even by a child. Moreover, instead of using a single large pendulum, which might be too massive, several smaller pendulums could be employed, producing the same or even greater effect. It would not be difficult to devise a method by which this kind of mechanism could be used without any inconvenience to navigation.

This section suggests that Bernoulli's pendulum propulsion mechanism could replace traditional rowing and might be especially useful in situations where wind is lacking. It also highlights the minimal effort required to maintain the system and the potential for using multiple smaller pendulums.

§.9. However, the utility of this method in navigation is too great to believe that it could have remained hidden for so long, especially since it does not involve a particularly obscure mechanism. Indeed, due to the very magnitude of its potential benefits, it justifiably arouses suspicion. This suspicion is further heightened by the

fact that the description of this device is only found in the posthumous works of Jacob Bernoulli, and was never published during his lifetime. It seems highly improbable that such a great man, now deceased, would have concealed such an invention—one that would certainly surpass all of his other inventions, even the greatest—unless he himself had doubts about its success. Therefore, if I were to demonstrate that no motion at all can be imparted to the ship by the impacts of such a pendulum, this would in no way detract from the praise and merits of this most eminent man, since he took great care during his life not to make public a theory that had not yet been perfected.

This section raises doubts about the practicality of Bernoulli's pendulum propulsion system, suggesting that the fact that it was not published during his lifetime indicates Bernoulli himself may have had doubts about its success. The author also emphasizes that disproving the mechanism would not diminish Bernoulli's reputation, as he was cautious about releasing unproven ideas.

§.10. If we wish to investigate the effect of such pendulum impacts, we must first determine how much the ship is pushed backward while the pendulum descends through the quadrant **BMC**. Then, the impact itself must be considered—how the ship is propelled forward by it, and the exact amount of motion imparted to the ship toward the bow must be determined. Finally, since this forward motion is again slowed by the subsequent ascent of the pendulum after its rebound, it will be necessary to examine whether the ship, after the pendulum has returned to point **B**, still has any residual forward motion, and if so, how much. For if the ship, having been at rest at the start of the descent, returns to a state of rest after the pendulum's ascent, and thus is again at rest at the beginning of the second descent, there will be no doubt that the ship will remain almost stationary in the same location, with the entire effect of the pendulum being consumed in alternating forward and backward motions that exactly cancel each other out. The determination of this reciprocal motion, if we wish to account for the resistance of the water, would be extremely difficult and could not be solved without a very laborious calculation.

This section outlines the process of analyzing the pendulum’s effect on the ship, emphasizing the challenge of accounting for water resistance and the possibility that the ship's motion could cancel out, leaving it stationary.

§.11. For this reason, I will endeavor to pursue another, simpler method by which the effect arising from successive impacts of such a pendulum can be just as clearly understood and distinguished. Specifically, I will consider the ship as being completely fixed in place and investigate the sum of the instantaneous forces by which the ship is pushed backward during each descent and ascent of the pendulum. Then, in a similar manner, I will separately express the force of the impact by which the ship would be propelled forward, so that in this way, the total force pushing the ship backward from each action of the pendulum and the propelling force can be known. Each action of the pendulum consists of three parts: the descent, the impact, and the ascent. If, therefore, the total of the forces pushing the ship backward from the descent and subsequent ascent is equal to the force of the impact propelling the ship forward, we will be able to safely conclude that even if the ship were free, no forward

motion would be induced. However, if either the force of the impact or the sum of the backward-pulling forces prevails, the free ship will either be pushed forward or backward, respectively.

This section describes the author's intention to analyze the pendulum's effects on a fixed ship by calculating the forces exerted in both directions (backward and forward) to determine whether, if the ship were free, it would move forward, backward, or remain stationary.

§.12. Since the ship is pushed toward the stern at every moment of the pendulum’s descent, we must determine the magnitude of this force for any given position of the pendulum **AM**, and multiply it by the infinitesimal element of time. This differential expression is then integrated, and when applied to the entire descent, it will yield the total sum of all the forces pulling the ship backward. In a similar manner, the total sum of the forces for the ascent is collected. Now, if the ship were free to respond to the action of these forces, it would acquire motion from them, the quantity of which—defined as the product of the ship’s mass and the velocity generated—would be exactly equal to the integral. Next, we must determine the quantity of motion that would be imparted to the ship by the pendulum’s impact if the ship were free, and compare this with the previously calculated quantity to see whether one is greater or if they are equal. In this way, we will be able to conclude with certainty whether the ship will acquire any motion from these forces or not.

This section outlines the mathematical approach for determining whether the ship would gain motion from the pendulum’s forces by calculating and comparing the total backward forces during the descent and ascent with the forward force from the pendulum’s impact.

§.13. Since in this investigation it is of great importance whether the pendulum is simple or compound, let us first assume that the pendulum is simple, so that its entire mass can be considered concentrated at its center of gravity **M**. Let this pendulum, in each ascent and descent, describe a complete quadrant **BMC**. Let the length of the pendulum be $\mathbf{AM = AC = a}$, and its weight be **M**. After descending from **B**, at time **t**, it has now reached the position **AM**, where it still deviates from the vertical line **AC** by the angle $\mathbf{CAM = \phi}$. The velocity at point **M** corresponds to the height $\mathbf{LM = a\cos\phi}$, hence the velocity at **M** is $\mathbf{v = \sqrt{(a\cos\phi)}}$. Thus, for an infinitesimal time **dt**, the pendulum sweeps through the arc $\mathbf{-ad\phi}$, and $\mathbf{dt = -ad\phi / \sqrt{(a\cos\phi)}}$. I will consistently follow this rule, expressing velocities through the square roots of their corresponding heights and the elements of time as the spaces traversed during the intervals, applied to the velocities.

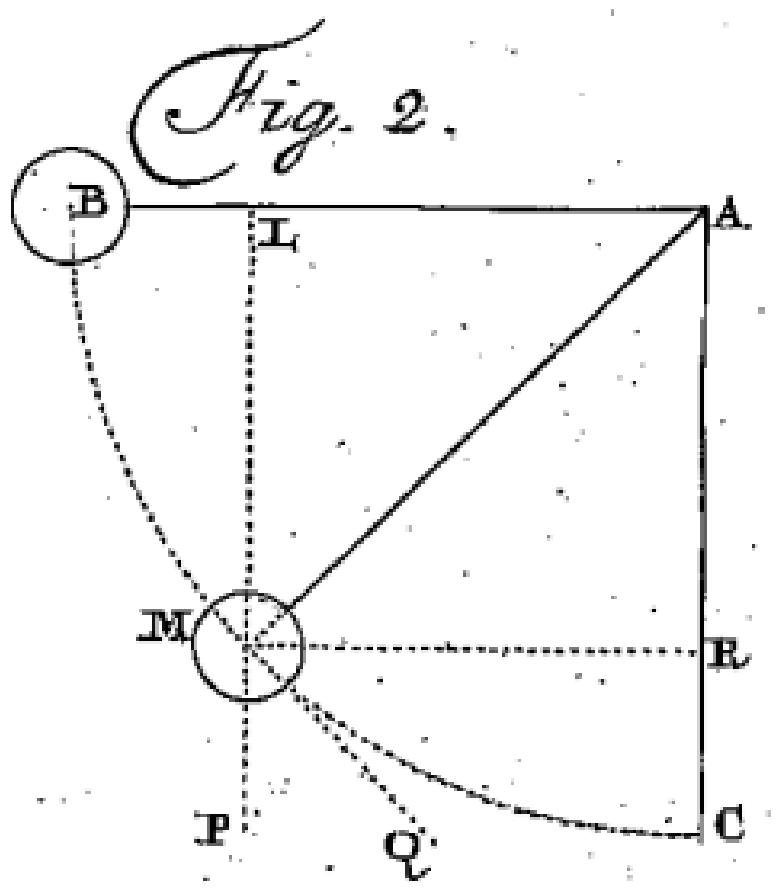

This section sets up the mathematical framework for analyzing the motion of a simple pendulum, introducing the key relationships between angle, velocity, and time.

§.14. Having determined the height corresponding to the velocity of the pendulum at **M** as **acosϕ**, the centrifugal force will be **2M acosϕ/a = 2M cosϕ**, directed along **AM**. Next, since the pendulum is pulled downward by gravity with a force of **M** in the vertical direction **MP**, this force can be resolved into two components: one along the normal to **AM** (**MQ**), giving **M sinϕ**, and another along **MA**[1], giving **M cosϕ**. The component **M sinϕ** is entirely used to accelerate the motion of the pendulum, so it does not affect the tension in the string **AM**. On the other hand, the component **M cosϕ** acts entirely in stretching the string **AM**. Therefore, the total tension in the string **AM** is the sum of the centrifugal force and the vertical component, giving **T = 3M cosϕ**. This tension pulls the suspension point **A** along the direction **AM**. From this, we resolve the force acting on the ship backward as **F = 3M cosϕ sinϕ**.

This section calculates the forces acting on the pendulum and the ship, focusing on the centrifugal force and gravitational components, which combine to produce a backward force on the ship.

§.15. Let this force, **3M cosϕ sinϕ**, which pushes the ship backward toward the stern, be multiplied by the element of time **dt = -adϕ / √(acosϕ) = -dϕ √(acosϕ)/cosϕ**, giving the instantaneous solicitation **-3M dϕ sinϕ √(acosϕ)**, which is equal to the generated element of motion. Since **-dϕ = sinϕ dcosϕ**, if we assume **cosϕ = z**, the instantaneous solicitation **3Mdz √(az)**. The integral of this expression gives the total solicitation, **2Mz √az** = **2M cosϕ √(a cosϕ)**. This expression provides the sum of all the forces by which the ship is pushed backward as the pendulum descends through the arc **BM**. Let **ϕ = 0**, and the total instantaneous solicitation from the complete descent of the pendulum is **2M√a**, which is equal to the sum of the forces resulting from the subsequent ascent. In each action of the pendulum, the ship is pushed backward by forces whose total is **4M√a**, and if the ship were free, the motion imparted would have a quantity of **4M√a**.

This section calculates the total force pushing the ship backward during the pendulum's descent and ascent, as well as the resulting motion if the ship were free.

[1] **Note**: originally **MR**, which has been corrected to **MA**.

§.16. Let us now also inquire how much force the pendulum exerts on the ship when it strikes the elastic platform **AF**, where we will consider the platform as immobile. The pendulum, with mass or weight **M**, strikes this platform with the velocity corresponding to the height **a**, from which it descended. To more clearly examine the effect of the collision, let us suppose that an elastic element **CD** is attached to the platform at point **C**, and the pendulum strikes this elastic element. The length of the elastic element can be conceived as arbitrarily small. At time **t**, from the moment the collision begins at point **D**, the pendulum reaches point **M**, and the elastic element is compressed into the state **MC.** Let the distance **DM = a**, and the residual velocity of the pendulum at point **M** corresponds to the height **v**, and the force of the elastic element **CM**, which attempts to expand back, is called **P**.

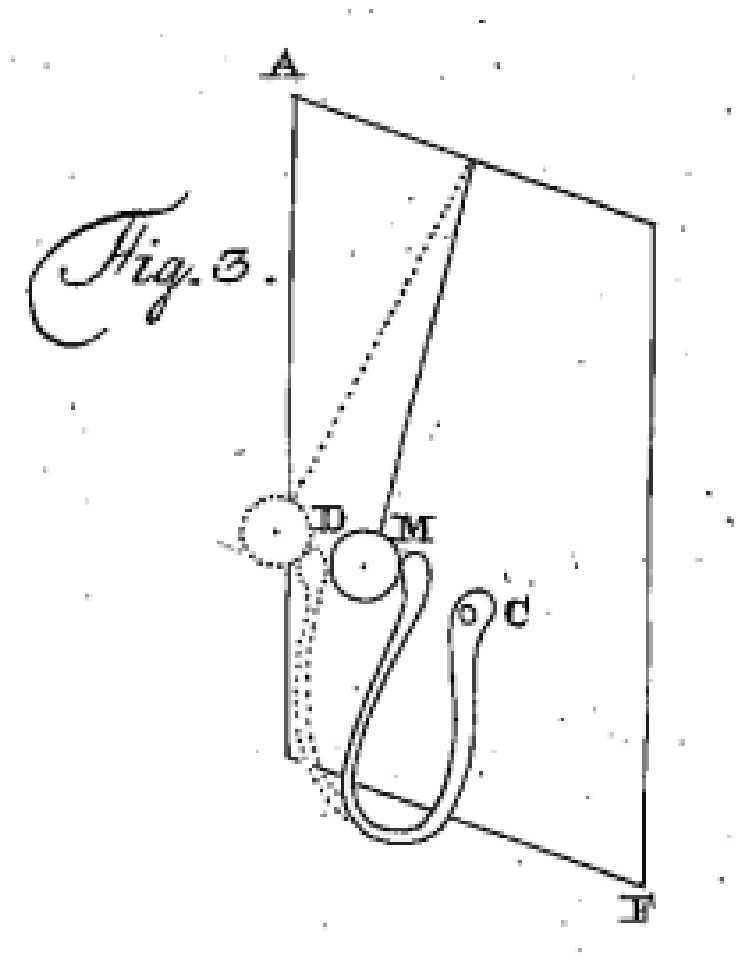

This section begins the analysis of the force exerted by the pendulum when it collides with an elastic platform, including a consideration of the elastic deformation during the collision.

§.17. With these conditions established, as the pendulum penetrates further by the small distance **$dx$**, by the laws of motion we have **$M\,dv = -P\,dx$**. But since the platform **AF**, and thus the ship itself, is being propelled forward by the force **P**, we must investigate the value of **$\int P\,dt$** during the time the collision lasts. Since **$dt = dx / \sqrt{v}$**, the previous equation becomes **$M\,dv / \sqrt{v} = -P\,dt$**, which simplifies to **$\int P\,dt = -\int (M\,dv / \sqrt{v}) = C - 2M\sqrt{v}$**, and since this quantity must vanish at the start of the collision, we have **$C = 2M\sqrt{a}$**, so **$\int P\,dt = 2M\sqrt{a} - 2M\sqrt{v}$**. Now, assuming both bodies are perfectly elastic, after the collision the pendulum will have a velocity equal to the velocity with which it struck, **$\sqrt{a}$**, but in the opposite direction, hence **$\sqrt{v} = -\sqrt{a}$**. Substituting this value, we find that the total instantaneous force arising from the collision and propelling the ship is **$P = 4M\sqrt{a}$**.

This section calculates the force exerted by the pendulum on the ship during the collision, assuming perfect elasticity.

§.18. The motion that the pendulum's impact attempts to impart to the ship toward the bow is exactly equal to the motion that the forces acting on the pendulum, while it completes one descent and ascent, are capable of generating in the opposite

direction. From this, it is evident that, even if the ship receives propulsion toward the bow from the impact of the pendulum, this entire motion will subsequently be completely nullified by the pendulum's ascent and subsequent descent. Since this destruction of motion occurs after each impact, the ship cannot achieve any progressive motion, as the renowned Jacob Bernoulli suspected. Although he employed nearly the same reasoning here and similarly estimated the forces pulling the ship backward, he made a certain error in determining the propelling force arising from the impact, which his commentator, the distinguished Cramer, clearly noticed but did not correct due to the difficulty of the necessary calculations.

This section concludes that the motion generated by the pendulum's impact is countered by the forces during the pendulum's ascent and subsequent descent, resulting in no progressive movement of the ship. It also mentions an error Bernoulli made, which was identified by his commentator, Cramer.

§.19. This perfect compensation between the propelling and repelling forces does not only occur when the pendulum moves through an entire quadrant; it is also observed when the pendulum oscillates through smaller arcs, and it will be worthwhile to demonstrate this. Let the simple pendulum **AM**, previously considered, descend through an arc **HMC** smaller than a quadrant, and let the pendulum length **AM = a**, the weight of the body **M = M**, and the angle **HAC = θ**. After the time **t** has passed and the pendulum has described the arc **HM**, let the angle **MAC = ϕ**. Drawing the horizontal lines **HI** and **MK**, the height **AI = a cos θ** and **AK = a cos ϕ**. Therefore, **IK = a(cos ϕ - cos θ)**, which is the height corresponding to the velocity of the body at **M**. Thus, the centrifugal force will be **2M (cos ϕ - cos θ)**, which stretches the string **AM**. As the pendulum descends through the arc **-adϕ** in the time **dt**, with velocity **√a(cos ϕ - cos θ)**, we have **dt = -dϕ √a/ √( cos ϕ - cos θ)**.

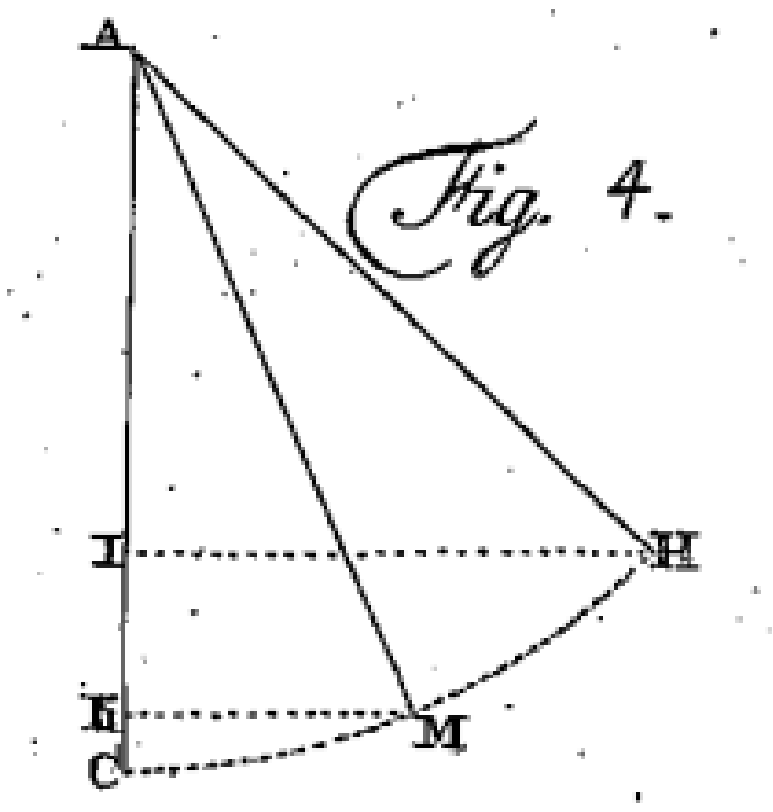

This section discusses the forces acting on the pendulum as it oscillates through an arc smaller than a quadrant, focusing on the changes in height and the resulting centrifugal force.

§.20. Let us now also consider the force of gravity, which urges the pendulum at point **M** downward along **MP** with a force of **M**. By resolving this force, we obtain the component which tends to stretch the pendulum with a force **MR = M cos ϕ**. Therefore, the string **AM** is altogether stretched by a force of **3M cos ϕ -**

**$2M \cos \theta$**. Since this force has an oblique direction, its horizontal component is **$3M \cos \phi \sin \phi - 2M \cos \theta \sin \phi$**. Thus, multiplying this by the element of time **$dt = -d\phi \sqrt{a} / \sqrt{(\cos \phi - \cos \theta)}$**, we get the instantaneous solicitation **$-M\, d\phi \sin \phi\, (3 \cos \phi - 2 \cos\theta)\sqrt{a} / \sqrt{(\cos \phi - \cos \theta)}$**. Let us set, and **$\cos \phi = z$**, and **$\cos \theta = b$**, so that **$-d\phi \sin \phi = dz$**, and the instantaneous solicitation will be **$M dz(3z-2b) \sqrt{a} / \sqrt{(z - b)}$**. Integrating this, we have: **$+2Mz\sqrt{a(z-b)} = +2M \cos \phi\sqrt{a (\cos \phi - \cos \theta)}$**, which should vanish at the initial time where **$\phi= \theta$**, thus giving **$C = 0$**. Therefore, the sum of all the instantaneous forces corresponding to the descent through the arc **HM** is **$2M \cos \phi \sqrt{a(\cos \phi - \cos \theta)}$**.

In this section, the author examines the gravitational force acting on the pendulum and how it contributes to the tension in the pendulum's string. By resolving the forces, the horizontal component that acts to pull the ship backward is determined.

§.21. Let us now set **$\phi=0$**, so that for the entire descent of the pendulum, the sum of the instantaneous forces will be **$2M\sqrt{a(1 - \cos \theta)} = 2M\sqrt{CI}$**; where **$\sqrt{CI}$** represents the velocity of the pendulum at the lowest point **C**. Thus, this sum will be equal to twice the quantity of motion that the pendulum acquires at point **C**. Since the ascent is similar to the descent, the total forces pushing the ship backward, arising from both the ascent and the descent, will be **$4M\sqrt{CI}$**. From §.17, we obtain the force resulting from the impact if, in place of the velocity **$\sqrt{a}$** considered there, we substitute the velocity with which the pendulum strikes the platform, which is **$\sqrt{CI}$**. By doing this, we find that the force arising from the collision is also **$= 4M\sqrt{CI}$**. Therefore, in this case, the forces pushing the ship backward from the descent and ascent, taken together, are equal to the force with which the ship is propelled forward by the impact. Hence, even in this case, the impacts of the pendulum cannot induce any progressive motion in the ship.

This section continues the analysis by considering the pendulum swinging through a full descent (setting **$\phi=0$**). It shows that the total backward forces acting on the ship during the pendulum's descent and ascent exactly equal the forward force resulting from the pendulum's impact with the platform. Consequently, the net effect is zero, and the ship cannot achieve any forward motion from the pendulum's impacts.

§.22. What has been demonstrated thus far concerning simple pendulums appears so closely connected with a certain most constant law of nature that we can now assert with certainty that the same perfect equality between the propelling and repelling forces will be found in any composite pendulums as well. Although this could easily be shown from the nature of the center of oscillation, it seems so consistent with the other laws of nature that it rightly deserves to be counted among the primary principles of mechanics. Just as in pressures (or so-called *dead forces*) the reaction is always equal and opposite to the action, so too in collisions a similar equality holds—which is all the less surprising since any collision can quickly be reduced to pressures. Therefore, any impact cannot produce more force than is required to generate motion in the colliding bodies; and from this it follows that ships cannot be propelled in this Bernoullian manner. Moreover, any other mechanisms that are entirely enclosed within the ship and rely on no external principle will be equally useless and incapable of imparting any motion to the ships.

This section concludes that the equality between the forces propelling and repelling the ship holds not only for simple pendulums but also for any composite pendulums. The author emphasizes that this principle aligns with fundamental laws of nature and mechanics. Consequently, any attempt to propel a ship using internal mechanisms without external interaction—such as Bernoulli's proposed pendulum system—will be ineffective.

§.24. The rationale of this principle is perceived much more clearly if we first assume water entirely devoid of resistance, so that the ship can perpetually continue any impressed motion without any impediment. In this hypothesis, if a pendulum or any other kind of machine is operated on the ship—which does not receive any external source of motion—it is manifest from the laws of motion that the common center of gravity of the ship and the machine must remain at rest, except insofar as it ascends or descends vertically. For this law is observed not only when the machine acts upon the ship through pressures, in which case equal forces are exerted both on the ship and on the machine, but also if impacts or collisions occur; the state of the center of gravity will not be otherwise disturbed. Therefore, however the machine existing within the ship is constructed, and whether its action is composed of pressures or impacts, the common center of gravity cannot acquire any motion along the horizontal plane. Consequently, no such machine will be suitable for propelling the ship[A].

This passage explains that, even in an idealized scenario where water offers no resistance, any internal mechanism (like a pendulum or other machine) that doesn't interact with an external reference point cannot cause the ship to move horizontally. This is because the internal forces produce equal and opposite reactions, leaving the overall center of gravity stationary in the horizontal direction. Thus, such machines cannot propel the ship forward.

§.25. But if the resistance of the water is also considered, then the previously mentioned law about the center of gravity is somewhat infringed, since the ship, when acted upon by the machine, does not yield as much as it ought to yield according to that law; and similarly, in collisions, due to the resistance of the water, the common center of gravity does not remain perfectly at rest. A very difficult calculation would be required if one wished to analyze each of these effects according to the precepts of mechanics. However, since the entire effect of the resistance is consumed in diminishing motion, and no motion can be produced by it, the resistance of the water certainly cannot be the cause by which motion is imparted to the ship, since the same ship, with the resistance removed, ought to remain at rest. Hence, we conclude with the utmost certainty that, just as a ship, with the resistance of the water removed, cannot acquire any progressive motion from internal forces, so much less can any motion be imparted by such forces if the resistance of the water is added.

This passage explains that even when considering water resistance, internal mechanisms within the ship cannot generate forward motion. The resistance of the water only serves to diminish motion, not create it. Therefore, if a ship cannot be propelled by internal forces in the absence of water resistance, it certainly cannot be propelled when water resistance is present.

§.26. Although this reasoning seems beyond all exception, yet there are cases in which, on account of resistance itself, motion is produced, when none would arise if it were removed. For if the ship **DEFG** rests on its base **EF** upon a rough plane, over which it cannot be moved without noticeable friction, it is clear that the friction can be so great that it cannot be overcome by the forces tending the pendulum, and thus no backward motion is impressed on the ship by these forces. Nevertheless, the friction can be overcome by the impact of the pendulum against the platform **AF**, whereby the ship will be pushed forward somewhat by each of the pendulum's impacts; and since this forward motion is not destroyed by contrary forces, the ship will indeed be propelled—a result that would in no way be obtained if there were no friction present. Herein lies a remarkable mechanical paradox: that friction itself can be the cause of some motion, so that if the friction were removed, no motion at all would follow[B].

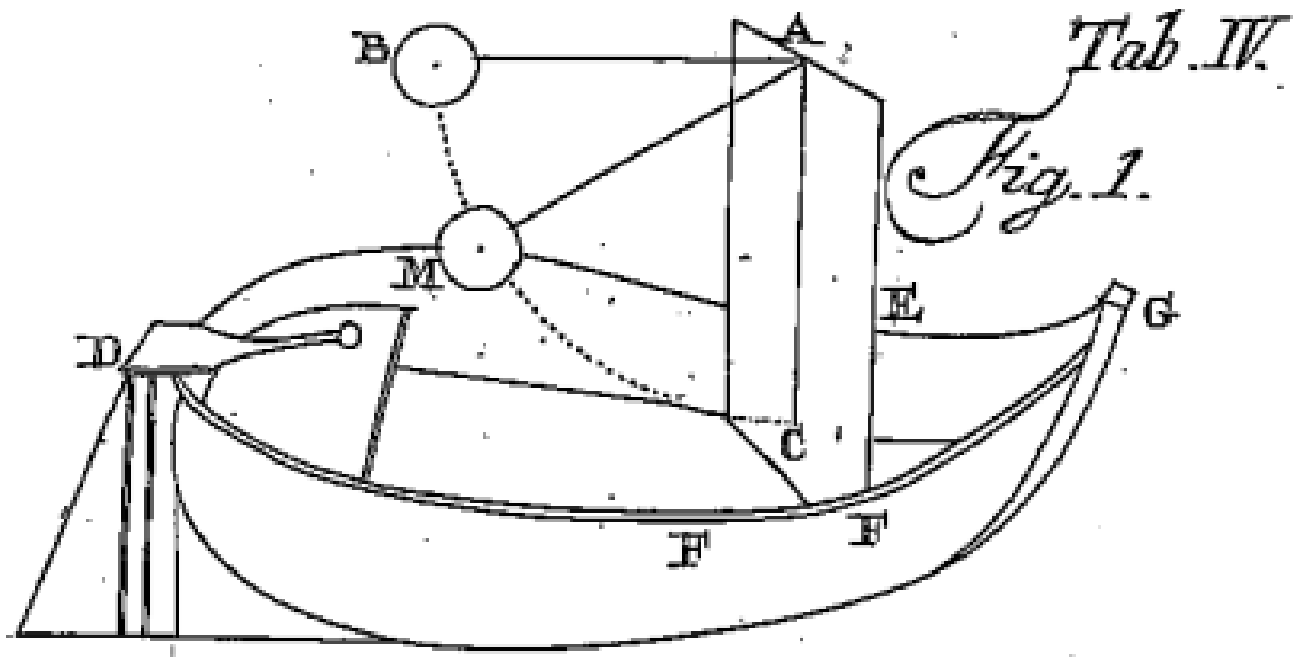

This passage discusses a paradox in mechanics where friction, typically considered a force that opposes motion, actually becomes the cause of motion under certain conditions. Specifically, when the ship rests on a rough surface with significant friction, the backward forces from the pendulum are insufficient to move the ship backward. However, the impacts from the pendulum can overcome this friction and push the ship forward. Since the backward forces are nullified by friction and the forward forces are not opposed, the ship gains forward motion—a situation that wouldn't occur without friction.

§.27. Therefore, an even greater reason for doubt arises from this, namely, whether due to the resistance of the water, any motion can be induced in the ship by impacts of such a pendulum, even if it is certain that, if the resistance were absent, no motion could be impressed upon it in this way. To resolve this doubt, let us consider a ship being acted upon alternately by two forces, **p** and **P**, one of which, **p**, urges it toward the prow for a time **t**, and the other, **P**, urges it toward the stern for a time **T**. These forces **p** and **P** are so proportioned to the times **t** and **T** that **pt = PT**, an equality provided by the previous determination of both the propelling and repelling forces. Although neither force **p** nor **P**, during the time each acts, has been found to be constant, nevertheless, for the convenience of calculation, we can assume both to be constant without error, since a slight inequality cannot be the cause of any motion that would not equally follow from equality.

In this passage, Euler addresses the question of whether water resistance allows the ship to gain any motion from the pendulum's impacts, even though it's known that without resistance, no motion would be imparted. By analyzing the ship being alternately pushed forward and backward by forces **p** and **P**, and setting their products with time equal (**pt = PT**), the author simplifies the problem to determine if any net motion results from these alternating forces.

§.28. Let us therefore suppose that the force **p** acts first, by which the ship is propelled, and that initially the ship was at point **A**, where it had a velocity toward the prow equal to **√b** , and has now covered the distance **AP = x**, and at point **P** has a velocity corresponding to the height **√v**. Since the resistance is proportional to the square of the velocity, let it be set as = **v/k**; thus, we have: **dv = pdx – vdx/k**. Let the time in which it goes from **A** to **P** be **t**; then **dt=dx/√b**. Substituting this value for **dx**, we will have: **kdv = (kp - v)dt√v**. Let **√b = c** and **√v = u** so that the irrationality is removed; it will be: **2kdu = (kp – uu)dt**. Since when **t = 0** it is **u = c**, the integral of this equation, although it could be expressed using logarithms, will be more conveniently expressed through a series in the following manner:

$$u = c + At + Btt + Ct^3 + Dt^4 + \text{etc.}$$

from which it follows:

$$\frac{2kdu}{dt} = 2Ak + 4Bkt + 6Cktt + \alpha Dkt^3 + \text{etc.}$$

$$kp - uu = kp - 2Act - 2Bctt - 2Cct^3 \; \text{etc.}$$
$$\qquad\qquad - cc \qquad - AAtt - 2ABt^3$$

The equation of coefficients will therefore give:

$$A = \tfrac{1}{2}p - \frac{cc}{2k}.\ B = \frac{-cp}{4k} + \frac{c^3}{4k^2}:$$

$$6Ck = \frac{ccp}{2k} - \frac{c^4}{2kk} - \tfrac{1}{4}pp + \frac{ccp}{2k} - \frac{c^4}{4kk} = -\tfrac{1}{4}pp + \frac{ccp}{k} - \frac{3c^4}{4kk}$$

Then, $C = \frac{-pp}{24k} + \frac{ccp}{6kk} - \frac{c^4}{8k^3}$

$$8Dk = \frac{cpp}{12k} - \frac{c^3p}{3kk} + \frac{c^5}{4k^3} + \frac{cpp}{4k} - \frac{c^3p}{2kk} + \frac{c^5}{4k^3}$$

or $D = \frac{cpp}{24kk} - \frac{5c^3p}{48k^3} + \frac{c^5}{16k^4}$ etc.

From these, therefore, arises the sought velocity of the ship at the finite time **t**:

$$u = c + \tfrac{1}{2}t\left(p - \frac{cc}{k}\right) - \frac{ctt}{4k}\left(p - \frac{cc}{k}\right) - \frac{t^3}{24k}\left(pp - \frac{4ccp}{k} + \frac{3c^4}{kk}\right)$$
$$+ \frac{ct^4}{48kk}\left(2pp - \frac{5ccp}{k} + \frac{3c^4}{kk}\right) + \text{etc.}$$

This passage sets up and begins to solve a differential equation related to the motion of a ship experiencing resistance proportional to the square of its velocity. The author chooses to solve the equation by expressing the solution as a power series expansion rather than using logarithmic expressions, which may simplify the calculations.

Fig. 5.

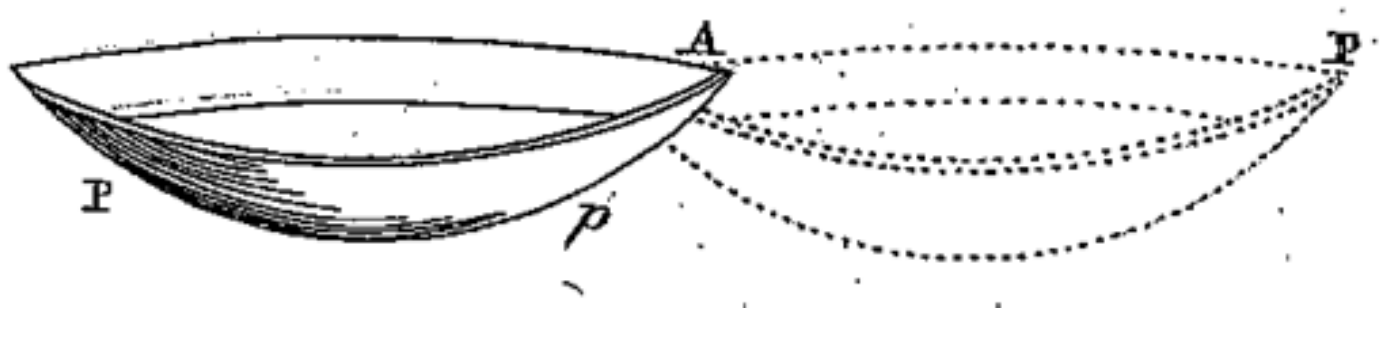

§.29. Similarly, if at the end of this time **t**, the forward velocity of the ship is assumed to be **C** so that **C = u**, and then the force **P** pulls the ship backward for a time **T**, if after this elapsed time **T**, the residual velocity of the ship is set to be **U**, it will be found that

$$U = C - \tfrac{1}{2}T\left(P + \frac{CC}{k} + \frac{CTT}{4k}\left(P + \frac{CC}{k}\right) - \frac{T^3}{24k}\right.$$
$$\left(PP + \frac{4CCP}{k} + \frac{7C^4}{kk} + \frac{CT^4}{48kk}\left(2PP + \frac{5CCP}{k} + \frac{3C^4}{kk}\right) - \text{etc.}\right.$$

Since indeed **pt = PT**, we set **pt = PT = Q**, so that **p = Q/t** and **P = Q/T**, substituting these values in place of **p** and **P**, we obtain:

$$u = c + \tfrac{1}{2}Q - \frac{cct}{2k} - \frac{cQt}{4k} - \frac{QQt}{24k} + \frac{c^3tt}{4kk} + \frac{ccQtt}{6kk} + \frac{cQQtt}{24kk} - \text{etc.}$$
$$U = C - \tfrac{1}{2}Q - \frac{CCT}{2k} + \frac{CQT}{4k} - \frac{QQT}{24k} + \frac{C^3T^2}{4kk} - \frac{CCQT^2}{6kk} + \frac{CQQT^2}{24kk}\ \text{etc.}$$

Moreover, since the time of impact **t** is considered almost infinitely small, with **t** approaching zero, it will be $\mathbf{u} = \mathbf{c} + \frac{1}{2}\mathbf{Q} = \mathbf{C}$, whence, from the subsequent action of the pendulum arising from its tension, it will result:

$U = c - \frac{T}{24k}(12cc + 6cQ + QQ) + \text{etc.}$

where we neglect the remaining terms because compared to these two, they are very small.

§.30. Hence, it is evident that **U < c**, and therefore, the velocity of the ship must decrease due to any action of the pendulum, which is composed first of the impact and then of the tension of the pendulum. Thus, even if the ship already has some forward-directed velocity, it will soon lose it due to the action of the pendulum; whence, much less can it acquire any motion from the pendulum when it is at rest. But if it is objected that the ship might perhaps be driven backward by the pendulum by exchanging the velocities **u** and **U**, it will be shown in a similar manner that any backward-directed velocity the ship might have will also be continually diminished by the action of the pendulum, and thus, in no way can any motion be imparted to the ship by such a pendulum.

---

[A] **Note:**

It was asked to ChatGPT to make a distinction between pressures and impacts. CHATGPT reply: In the context of the passage you've provided, the distinction between **pressures** and **impacts** refers to two different types of forces or interactions within mechanical systems, especially concerning how they influence motion.

**Pressures** (Latin: *pressiones*)
**Definition**: Pressures are continuous, steady forces applied over a period of time without causing relative motion at the point of contact. They involve sustained contact between objects.
**Characteristics**:

- **Static Interaction**: Pressures are associated with static or slowly changing situations where forces are applied gradually.
- **Dead Forces** (Latin: *vires mortuae*): Historically, pressures were linked to "dead forces," which are forces resulting from potential energy or static pressure. These forces do not involve actual motion but have the potential to cause motion if unopposed.

**Examples**:

- Pushing against a wall.
- The force exerted by a compressed spring that is held in place.
- The weight of an object resting on a surface.

**Impacts** (Latin: *percussiones*)
**Definition**: Impacts are sudden, transient forces resulting from collisions or abrupt interactions between bodies. They involve a rapid exchange of momentum over a very short time.
**Characteristics**:

- **Dynamic Interaction:** Impacts are associated with dynamic situations involving motion and sudden force application.
- **Living Forces** (Latin: *vires vivae*): Historically, impacts were associated with "living forces," which involve kinetic energy and actual movement. These forces are capable of doing work due to the object's motion.

**Examples:**

- A hammer striking a nail.
- A billiard ball colliding with another ball.
- A pendulum bob hitting a barrier.

**Context in the Passage**
In the passage, the author discusses the inability of internal mechanisms to propel a ship forward when operating solely through internal forces, whether those forces are pressures or impacts.

- **Conservation of Momentum:** According to the laws of motion, particularly Newton's third law (for every action, there is an equal and opposite reaction), internal forces within a closed system cannot change the system's overall momentum. The center of gravity of the system remains stationary unless acted upon by an external force.
- **Application to Pressures:** When a machine within the ship exerts a continuous force (pressure) against another part of the ship, the action and reaction forces cancel out internally. The ship cannot gain net motion from these internal pressures because they do not interact with an external reference point (like the bottom or the banks of a channel).
- **Application to Impacts:** Even though impacts involve dynamic forces and momentum exchange, when they occur entirely within the ship, they also result in internal action-reaction pairs. The sudden forces from collisions between parts of the ship or its mechanisms cannot produce net movement of the ship as a whole.

**Underlying Principle**
The key principle here is the conservation of the center of mass motion in a closed system:

- **Closed System:** The ship and all mechanisms within it constitute a closed system when no external forces are acting.
- **Internal Forces:** Forces exchanged between parts of the system (whether pressures or impacts) are internal and cannot change the total momentum of the system.
- **External Forces Required:** To change the ship's momentum (i.e., to propel it forward), an external force must be applied. This could be interaction with the water (oars, propellers) or air (sails).

**Conclusion**

**No Net Propulsion from Internal Forces**

Both pressures and impacts, when confined within the ship and not involving any external interaction, cannot produce net propulsion. The internal forces cancel out due to action and reaction, leaving the ship's center of mass stationary in horizontal motion.

**Historical Context**

- **Dead vs. Living Forces:** The passage references a historical debate in physics regarding "dead" and "living" forces, associated with static potential energy and dynamic kinetic energy, respectively.
- The author asserts that regardless of this distinction, internal mechanisms relying solely on either type cannot propel the ship.

**Summary**

- **Pressures** are continuous, static forces (dead forces) applied internally, resulting in no net movement due to internal action-reaction cancellation.
- **Impacts** are sudden, dynamic forces (living forces) from collisions, but when occurring internally, they also cannot change the ship's momentum.
- **Implication:** Internal mechanisms, whether operating through pressures or impacts, cannot propel the ship without external interaction.

[B] **Note**:

This question was posed to ChatGPT: Since it has been proved that the forward force from the impact is equal to the backward force, for what reason then this backward force would not push the ship backward, since the force of the forward push has the same intensity? CHATGPT reply: That's an excellent and insightful question. The key to understanding this apparent paradox lies in the role that friction plays in the scenario described in §.26.

**Summary of the Situation**

- **Equal and Opposite Forces**: As established earlier in the text, the forward force exerted on the ship by the pendulum's impact is exactly equal in magnitude to the backward force exerted during the pendulum's descent and ascent due to internal tensions.
- **Ideal Conditions (No Friction)**: In an ideal, frictionless environment (either on water without resistance or on a smooth surface), these equal and opposite forces cancel each other out, resulting in no net movement of the ship. The internal forces obey Newton's third law, and the center of mass of the system (ship plus pendulum) remains stationary in the horizontal direction.

**Introduction of Friction:**

- **Ship on a Rough Surface**: When the ship is placed on a rough surface with significant friction between the ship's base and the ground, the dynamics change.
- **Effect on Backward Force**:

Static Friction Threshold: The backward force exerted on the ship during the pendulum's swing (descent and ascent) is spread out over a longer period and consists of continuous, smaller forces (tensions in the pendulum's support).

Insufficient to Overcome Friction: These backward forces are below the static friction threshold. The frictional force opposing the ship's movement is greater than the backward force trying to move it. As a result, the ship does not move backward; the backward forces are effectively neutralized by friction.

**Further Reflection:**

- This paradox highlights the importance of external interactions in mechanical systems. Internal forces alone cannot change the center of mass motion in a closed system, but external factors like friction can influence how internal forces manifest as motion.
- It also demonstrates the nuanced role of friction—not only as a force that opposes motion but also as one that can enable motion under certain conditions.